\documentclass[twocolumn]{aastex631}

\newcommand{\HI}{H$\,${\sc i} }
\newcommand{\HII}{H$\,${\sc ii} }

\usepackage{amsmath}
\usepackage{threeparttable}

\begin{document}

\title{A Long Stellar Stream in M83: Possible Connection Between XUV Disks and Minor Mergers?}

\correspondingauthor{Itsuki Ogami}
\email{ogami@tmu.ac.jp}

\author[0000-0001-8239-4549]{Itsuki Ogami}
\affiliation{National Astronomical Observatory of Japan, 2-21-1 Osawa, Mitaka, Tokyo 181-8588, Japan}
\affiliation{The Institute of Statistical Mathematics, 10-3 Midoricho, Tachikawa, Tokyo 190-8562, Japan}
\affiliation{Department of Advanced Sciences, Faculty of Science and Engineering, Hosei University, 3-7-2 Kajino-cho, Koganei, Tokyo 184-8584, Japan}

\author[0000-0002-7866-0514]{Sakurako Okamoto}
\affiliation{Subaru Telescope, National Astronomical Observatory of Japan, 650 North A’ohoku Place, Hilo, HI 96720, U.S.A.}
\affiliation{The Graduate University for Advanced Studies (SOKENDAI), 2-21-1 Osawa, Mitaka, Tokyo 181-8588, Japan}
\affiliation{National Astronomical Observatory of Japan, 2-21-1 Osawa, Mitaka, Tokyo 181-8588, Japan}

\author[0000-0001-7934-1278]{Annette M.N. Ferguson}
\affiliation{Institute for Astronomy, University of Edinburgh, Royal Observatory, Blackford Hill, Edinburgh, EH9 3HJ U.K.}

\author[0000-0002-3852-6329]{Yutaka Komiyama}
\affiliation{Department of Advanced Sciences, Faculty of Science and Engineering, Hosei University, 3-7-2 Kajino-cho, Koganei, Tokyo 184-8584, Japan}

\author[0000-0002-9053-860X]{Masashi Chiba}
\affiliation{Astronomical Institute, Tohoku University, Aoba-ku, Sendai, Miyagi 980-8578, Japan}

\author[0000-0002-8762-7863]{Jin Koda}
\affiliation{Department of Physics and Astronomy, Stony Brook University, Stony Brook, NY 11794-3800, U.S.A.}

\author[0000-0002-8758-8139]{Kohei Hayashi}
\affiliation{National Institute of Technology, Sendai College, Sendai, Miyagi 989-3128, Japan}
\affiliation{Astronomical Institute, Tohoku University, Aoba-ku, Sendai, Miyagi 980-8578, Japan}
\affiliation{Institute for Cosmic Ray Research, The University of Tokyo, Kashiwa, Chiba 277-8582, Japan}

\author[0009-0009-9769-534X]{Yoshihisa Suzuki}
\affiliation{Astronomical Institute, Tohoku University, Aoba-ku, Sendai, Miyagi 980-8578, Japan}

\begin{abstract}
We present the confirmation and characterization of a long stream (S-stream) in the southern part of M83. This feature is revealed using deep wide-field photometric data obtained by the Hyper Suprime-Cam (HSC) mounted on the Subaru Telescope. Using individual red giant branch (RGB)  stars, we successfully trace the stream over a large length of $\sim 81$~kpc and a considerable width of $\sim 9$ kpc. With a mean surface brightness of ${\langle \mu_{\it V} \rangle} \sim 31.8_{-1.9}^{+1.3}$ mag arcsec$^{-2}$, it is one of the most diffuse extragalactic streams currently known. The mean photometric metallicity of the stream is $\langle[{\rm M/H}]\rangle = -1.23\pm0.02$ dex with a standard deviation of $0.28\pm0.01$ dex, and we estimate the stellar mass to be $(8.5_{-2.8}^{+4.2}) \times 10^6~{\rm M_\odot}$ from the luminosity of RGB stars. Compared to its well-known northern counterpart, the S-stream is slightly more metal-poor, but our large-area RGB map shows compelling evidence that these two features are related, originating from a single low-mass merger event. 
The projected RGB distribution shows apparent density variations along the S-stream, although their significance and physical origin remain uncertain. Similarities between the morphology of the S-stream and some features in the \HI distribution suggest that a minor merger event may have disturbed and redistributed M83's outer \HI gas, leading to triggered star formation and the formation of the XUV disk. 

\end{abstract}
\keywords{Stellar streams (2166) --- Galaxies (573) -- Stellar Populations (1622) -- Photometry (1234)}

\section{INTRODUCTION}\label{section:intro}
The disk components of spiral galaxies hold critical information about their history of formation and evolution. Inside-out growth is commonly thought to be their main formation channel \citep[e.g.,][]{mo1998}. In this scenario, gas first accumulates in the central regions, where vigorous star formation is initiated. Over time, gas continues to accrete onto the outer parts of the galactic disk, facilitating star formation at larger galactocentric radii. This radial growth process is supported by observational evidence, such as radial gradients in stellar age and/or metallicity \citep[e.g.,][]{tinsley1978,munoz-mateos2007,wang2010}, which consistently indicate an outward progression of star formation over time.

Extended ultraviolet (XUV) disks are one of the notable examples of ongoing star formation in outer disk regions. XUV disks are characterized by ultraviolet (UV) emission that extends significantly beyond the optical radius ($R_{25}$; the radius corresponding to the isophote at a {\it B}-band surface brightness of 25 mag arcsec$^{-2}$.) of a galaxy \citep[e.g.,][]{thilker2007}, and their existence has been reported in numerous systems through UV imaging surveys, such as the Galaxy Evolution Explorer \citep[GALEX;][]{martin2005}. Since UV emission predominantly originates from young massive stars with short lifetimes, XUV disks provide compelling observational evidence of recent star formation in the outermost parts of galactic disks.

While the existence of XUV disks has been known for roughly twenty years, the physical mechanisms responsible for their formation remain a subject of debate. Several processes have been proposed, including external gas accretion \citep[e.g.,][]{thilker2007,lemonias2011}, radial transport of gas driven by internal structures such as bars and spiral arms \citep[e.g.,][]{bush2010,bresolin2016}, and gravitational interactions with companion galaxies \citep[e.g.,][]{bush2008}. However, it is still unclear which of these mechanisms plays the dominant role in XUV disk formation, or to what extent their contributions vary within individual galaxies.

Among the various scenarios proposed, mergers with satellite galaxies are considered to be one of the leading candidates. Since star formation in XUV disks occurs well beyond the optical radius, the presence of cold dense \HI gas is a necessary condition to initiate and sustain star formation in these outer regions. The supply and redistribution of \HI gas can be induced by minor mergers or gravitational interactions with companion galaxies. For instance, tidal interactions may draw gas from the inner disk toward the outskirts of the host galaxy or locally enhance the gas density in the outer regions. In some cases, the perturbation of the \HI disk by the accretion of satellites can compress gas and trigger localized star formation \citep[e.g.,][]{cox2008}. Furthermore, if the satellite itself contains a significant reservoir of \HI gas, it may act as a direct source of fuel for star formation upon merging \citep[e.g.,][]{bush2008}. Observational evidence supporting these mechanisms includes the detection of tidal-driven gas redistribution and resonance-induced density waves in the outer disk regions \citep{holwerda2012}. Such a picture is consistent with the hierarchical structure formation scenario predicted by the $\Lambda$-dominated cold dark matter cosmological model, wherein large spiral galaxies grow through the continuous accretion and merging of smaller systems \citep[e.g.,][]{white1991}. 


M83, which is also known as the Southern Pinwheel Galaxy, is a prototypical XUV disk which resides in the nearby Centaurus group.  Its prominent XUV disk was first discovered by  \citet{thilker2005} using GALEX observations, and \HII regions coincident with the XUV disk were later revealed by deep H$\alpha$ imaging with the Subaru telescope \citep{koda2012}. It is a system that has long been known for its extremely extended \HI disk \citep{huchtmeier1981,eibensteiner2024}, and more recently, there have also been detections of molecular gas \citep{koda2022,koda2024}. Moreover, many dwarf galaxies are known in the vicinity of M83 \citep[e.g.,][]{muller2015,muller2017}, and one stellar stream in the northern disk (hereafter, N-stream) was also detected \citep{malin1997}. \citet{barnes2014} estimated the total luminosity of the N-stream to be $1.9 \times 10^8~L_{\odot}$ and its total stellar mass as $1 \times 10^8~M_{\odot}$ using near-infrared imaging data from the Spitzer Space Telescope. 

Given the proximity of M83 at a distance of $\sim 4.7$ Mpc \citep{radburn-smith2011}, it is possible to undertake a search for very low surface brightness structures around it using resolved red giant branch (RGB) stars. This can shed light on M83's history of past interactions, and provide clues about the origin of its extended \HI and XUV disks.  M83 is the closest canonical XUV disk and one of the few XUV systems in which this kind of analysis can be undertaken. The Hyper Suprime-Cam (HSC), a prime focus camera on the Subaru Telescope with a FoV diameter of 1.5 degree \citep{miyazaki2012,miyazaki2018,furusawa2018,komiyama2018a}, is perfectly suited for a sensitive wide-area study of M83. In fact, several studies have already used HSC to survey nearby galaxy outer disks and stellar halos \citep[e.g.,][]{tanaka2017,okamoto2019,okamoto2023,zemaitis2023}. 



We present here the results of a study of the outskirts of M83 in which we detect a giant stellar stream in the southern part of the galaxy (hereafter, S-stream). As our paper was nearing completion,  \citet{bell2026} announced the discovery of this feature based on their joint analysis of a Subaru/HSC {\it g}/{\it r}/{\it i}-band dataset and Hubble Space Telescope observations. Our study provides independent confirmation of this feature, and characterizes its photometric and structural properties in detail. The paper is organized as follows. In Section \ref{section:data}, we present the details of the HSC data and the data reduction methods and calibration. We describe our detection of the S-stream in Section \ref{section:detection} and derive its properties in Section \ref{section:properties}. Finally, we discuss the implications of our findings and present our conclusions in Section \ref{section:conclusions}.


\section{OBSERVATION AND DATA REDUCTION}\label{section:data}

In this study, we use a combination of HSC datasets to investigate the outer regions of M83. The main dataset (\texttt{OBJECT} name is `NGC~5236'), which we analyze was acquired on the night of March 21, 2015 (PI: J. Koda; Proposal ID: S15A-017). This pointing was centered on M83, and was obtained in order to explore the XUV disk. The observations were conducted under average seeing conditions of $0\farcs78 \pm 0\farcs14$, with total exposure times of 10,530 seconds for the {\it g}-band and 3,570 seconds for the {\it i}-band.  In addition, we also exploit public data from the National Astronomical Observatory of Japan/Subaru-Mitaka-Okayama-Kiso Archive (SMOKA). In particular, we use the two M83 pointings (\texttt{OBJECT} names are `M83\_1' and `M83\_2') from Proposal ID: S15A-0281 (PI: E. Bell). These pointings include the main body of M83 but are centered to the west of it, extending into the halo region. These archival data were obtained on March 26 and 27, 2015. While the average seeing conditions in this dataset are very good overall ($0\farcs77 \pm 0\farcs13$ in each band), some individual exposures suffered from poor conditions. To ensure reliable star-galaxy separation at faint magnitudes, we opt to use only those images which satisfy seeing conditions of $< 1.1$ arcsec, yielding total integration times of 3,960 seconds in each of the {\it g}- and {\it i}-bands in the `M83\_1' field, and 3,960 seconds for the {\it g}-band and 3,060 seconds for the {\it i}-band in the `M83\_2'.  By combining these datasets, we improve the depth and spatial coverage of our imaging of M83's outer regions.

\citet{bell2026} analyzed the `M83\_1' and `M83\_2' pointings in the {\it g}, {\it r}, and {\it i} bands. Our imaging combines the archival {\it g}/{\it i}-band data from these fields with the M83-centered `NGC~5236' pointing described above. The additional pointing provides deeper imaging particularly in the eastern region of M83; the improvement in depth is therefore spatially dependent rather than uniform across the survey footprint. This difference is relevant when comparing the extent and properties of the S-stream measured in the two studies. The analyses also differ in their source-selection procedures: \citet{bell2026} adopted an HST-trained nearest-neighbor classifier for star--galaxy separation, whereas we use a determinant-radius selection calibrated using artificial-star tests on our hscPipe version 8.4 reductions. For their quantitative Subaru RGB metallicity and mass estimates, \citet{bell2026} selected stars brighter than $i_0=25.5$ and estimated contamination using distant background fields, whereas our RGB selection extends to $i_0=26$ and uses a rotated off-stream region to estimate contamination. Consequently, differences between the measurements may reflect both the spatially varying imaging depth and differences in the adopted footprints, source classification, RGB selection, contamination subtraction, and luminosity-to-mass or star-count-to-mass conversions.



The raw images are processed and calibrated using the HSC pipeline version 8.4 \citep[hscPipe;][]{bosch2018}, which is based on the software suite being developed for the Vera C. Rubin Observatory data \citep{ivezic2008,axelrod2010,juric2017}. Using this pipeline, the de-bias, dark subtraction, and flat-fielding corrections are conducted. After these steps, we follow \citet{zemaitis2023} and \citet{okamoto2024} by modelling and subtracting the sky component using a mesh of 32 pixels. Subsequently, photometric calibration is performed based on the Pan-STARRS 1 catalog \citep{magnier2013,chambers2016}, followed by stacking the individual exposures. Finally, the point spread function (PSF), cModel, and Kron model photometry are extracted from the stacked frames. 

In hscPipe, the \texttt{extendedness} parameter is provided to determine whether the object is a point source (\texttt{extendedness == 0}) or an extended source (\texttt{extendedness == 1}). This parameter is calculated as the ratio of the PSF flux to the cModel flux. The cModel in hscPipe \citep{bosch2018} is based on the SDSS cModel \citep{lupton2001,abazajian2004} and is a multiple PSF-convolved galaxy model. As it is designed to accurately reproduce the fluxes of galaxies, the combination of cModel and PSF fluxes serves as a powerful diagnostic tool for distinguishing between point-like and extended sources \citep{aihara2018}. However, cModel photometry is highly sensitive to even slight variations in the background level, and it often fails to yield reliable measurements, even in low-density regions. When this occurs, the \texttt{extendedness} parameter can take on \texttt{NaN} values, introducing artificial gaps in the source density across a field and significantly affecting the source completeness.

To avoid this problem, we choose to exclude extended sources using the \texttt{determinant radius} parameter, which was introduced by \citet{aihara2018}. Briefly, for each source, we compute the difference between the determinant radius, which represents the intrinsic size, and the PSF model width, in each band. Stars are expected to have values of this parameter near zero, whereas intrinsically extended sources will be offset to positive values. As described in Appendix \ref{section:detR}, we use artificial stars to determine the behaviour of this difference as a function of magnitude, and select stellar sources to be those objects that lie within 3$\sigma$ of the stellar locus in either the {\it g}-band or the {\it i}-band.

To evaluate our completeness and photometric quality, we perform artificial star tests using hscPipe and \texttt{injectStar.py}, following \citet{ogami2025}. In short, artificial stars with given magnitudes are embedded into the reduced stacked images at intervals of 300 pixels, using \texttt{injectStar.py}. About 250,000 artificial stars are injected into the images in intervals of 1 mag from 20 mag to 29 mag, in both the {\it g-} and {\it i} bands. We then rerun the detection and photometry using hscPipe. From these tests, we estimate that the 50\% detection completeness of the analyzed region is 26.29 mag in the {\it g}-band and 25.38 mag in the {\it i}-band. In addition, we estimate that the 5-sigma detection limits are 27.12 mag in {\it g}-band and 26.98 mag in {\it i}-band. These values depend on the crowding level, and the numbers we report here are valid over the region $-0.7 < \xi < 0.7$ and $-0.7 < \eta < 0.7$, where $(\xi,~ \eta)$ denote tangential coordinates centered at the M83 center, excluding the inner $ < 0.2$ degrees. The depths of the resulting photometric catalogs reach approximately 2 mag below the tip of the red giant branch (TRGB) at the distance of M83, which we assume throughout this paper to be 4.7 Mpc \citep{radburn-smith2011}.

The completeness limits should be distinguished from a direct comparison of imaging depth with \citet{bell2026}. Our estimates are based on artificial-star recovery using our hscPipe-based detection and photometry procedures, whereas \citet{bell2026} assess recovery and classification performance using HST reference stars. The resulting completeness estimates depend on the stellar populations and spatial regions sampled, as well as on the detection and source-selection procedures. The 50\% completeness limits ($i = 25.38$ mag) in this paper therefore cannot, by the value in \citet{bell2026} ($i = 25.5$ mag), quantify the relative imaging depth without matching these definitions and sampling regions. In particular, a completeness limit evaluated over a broad region need not capture the spatially localized improvement provided by our additional M83-centered pointing.

Using maps of the Galactic dust extinction \citep{schlegel1998,schlafly2012}, we apply an  extinction correction for each source. The value of $E({\it B}-{\it V})$ interpolated to each source position is extracted from the Python module \texttt{dustmaps}. The coefficient of the extinction-corrected magnitude in each band is obtained using the same method as in \citet{ogami2024}.  Briefly, we convolve the interstellar absorption curve of \citet{fitzpatrick1999} with $R_V = +3.1$ by the SED of a G-type star (${\it T}_{\rm eff} = 5000$ K, ${\it Z} = 0.1$, $\log{\it g} = 2.5$) and integrate it using the response curve of each filter. In the following sections, the subscript zero `$0$' means the extinction-corrected magnitudes.

\begin{figure*}[!ht]
 \begin{center}
  \includegraphics[width=180mm]{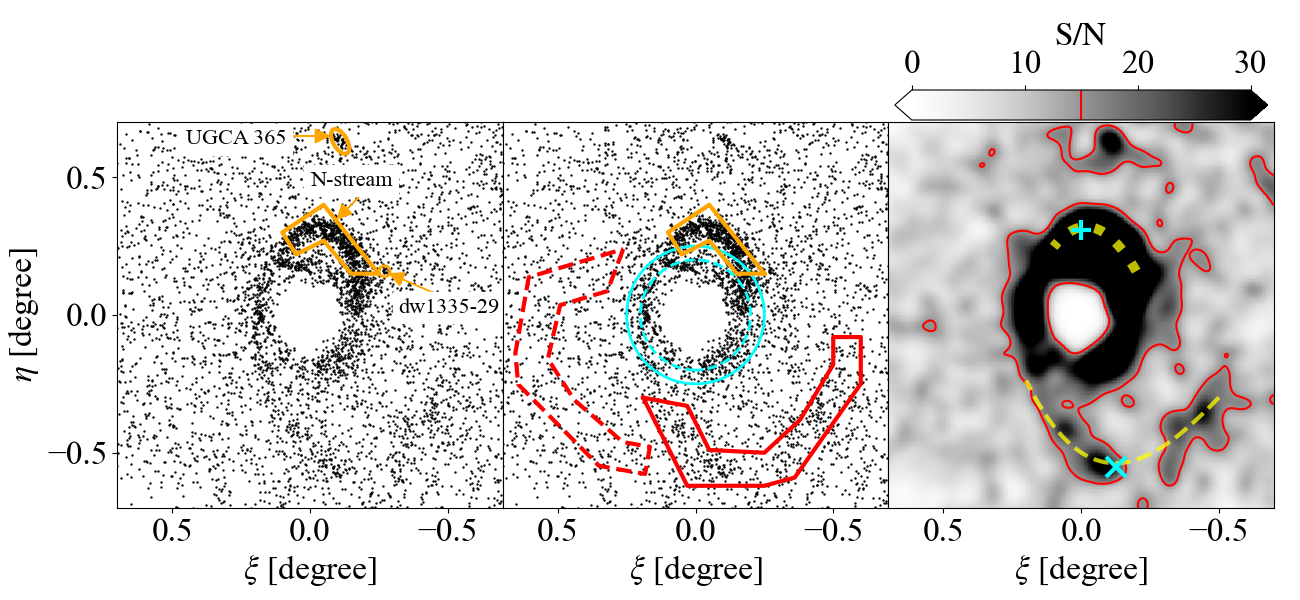}
 \end{center}
 \caption{Left: The spatial distribution of RGB stars around M83. These stars are extracted from the CMD box defined by the solid red polygon in Figure \ref{fig:CMD}. The known substructure and satellite galaxies are outlined in orange. Middle: The same as in the left panel, but with the solid red polygon indicating the recently-detected S-stream. The dashed red polygon shows the off-stream field, defined by rotating the solid red polygon 90 degrees clockwise around the center of M83. The cyan solid (dashed) circle indicates a region at 0.25 (0.20) degree from the center of M83. Right: the signal-to-noise ratio (S/N) map smoothed using a Gaussian kernel with a bandwidth of 1.2 arcmin. The red contour shows where the S/N reaches a value of 15. The yellow dashed and dotted lines represent the ridge lines of S-stream and N-stream, respectively. The cyan `x'-mark (`+'-mark) corresponds to the densest parts of the S-stream (N-stream). }\label{fig:SpatialDistribution}
\end{figure*}

\section{SPATIAL DISTRIBUTION OF THE SOUTHERN STREAM}\label{section:detection}

Figure \ref{fig:SpatialDistribution} shows the spatial distribution of RGB stars in the outer regions of M83. The RGB stars are selected as sources which lie within the solid red polygon shown in the color-magnitude diagram (CMD) in Figure \ref{fig:CMD}. This is defined as the area covered by PARSEC isochrones spanning [M/H] from $-1.51$ to $-0.95$, assuming the distance to M83 and an age of 10 Gyr. This range corresponds to the $1 \sigma$ uncertainty of the metallicity estimated for the stellar stream identified in this study, as described in Section \ref{subsection:metallicity}. The bright limit of the RGB box corresponds to the line connecting the TRGB of each isochrone, while the faint limit is set to $i_0=$26 mag, providing a good compromise between the 5-sigma detection and 50\% completeness limits (see Section \ref{section:data}).

In the left panel of Figure \ref{fig:SpatialDistribution}, the orange circle and ellipse highlight the locations of the known satellite galaxies, dw1335-29 at $(\xi,~\eta)\sim(-0.25,~0.2)$ and UGCA~365 at $(\xi,~\eta)\sim(-0.1,~0.6)$, both of which are clearly recovered in our RGB star-count map.   In addition, the orange polygon in the left and middle panels of Figure \ref{fig:SpatialDistribution} delineates the known N-stream, centered at $(\xi,~\eta)\sim(0,0.3)$.  This stream was identified by \citet{malin1997} and is known to continue down to around $(\xi,~\eta) \sim (0.2,0.05)$. Besides these known features, we also identify the southern stellar stream recently reported by \citet{bell2026} (hereafter, the S-stream). Our deeper and wider-area map provides independent confirmation of this feature and allows us to trace it further to the east, where the previous work could not trace it unambiguously \citep[see, Figure 6 in][]{bell2026}. As viewed in projection, we see the S-stream skirting the outer disc, and then smoothly connecting to the N-stream. The surface brightness of the S-stream decreases systematically with increasing projected distance from the main body of M83, and is everywhere lower in surface brightness than the N-stream.  Our findings strongly support the suggestion of \citet{bell2026} that these features could be connected, with the S-stream being the trailing arm of the N-stream. In the middle panel of Figure \ref{fig:SpatialDistribution}, the S-stream is outlined by a solid red polygon, the boundaries of which are determined based on the signal-to-noise ratio (S/N) calculations described below. We derive the stream ridge line by fitting a third-order polynomial to the enclosed RGB star positions, shown as the yellow dashed curve in the right panel of Figure \ref{fig:SpatialDistribution}. The length of the S-stream, as measured along the ridge line, is 0.99 degree ($\sim 81$ kpc).  The stream width (0.11 degree; $\sim 9$ kpc) is calculated from the standard deviation of stellar positions relative to the ridge line. Using the sources within the red polygon, we estimate that the average projected separation from the center of M83 to one of the densest regions along the S-stream (indicated by a cyan `x'-mark in the right panel of Figure \ref{fig:SpatialDistribution}) is approximately 0.57 degrees, which corresponds to 46 kpc at the distance of M83. We measure the properties of the N-stream in a similar manner, finding the length and width to be 0.40 degree ($\sim 33$ kpc) and 0.10 degree ($\sim 8$ kpc; shown as a cyan `+'-mark in the right panel of Figure \ref{fig:SpatialDistribution}), respectively.

To quantify the significance of the detection of the S-stream, we calculate the S/N of this stream according to the prescription of \citet{yang2023}. This requires a reference field devoid of known structures, located at the same distance from M83. We adopt the original red polygon rotated by 90 degrees clockwise as the off-stream field, which is shown as the dashed red polygon in the middle panel of Figure \ref{fig:SpatialDistribution}. Then, we apply a Gaussian kernel with a bandwidth of 1.2 arcmin to smooth the spatial distribution and calculate the S/N through the formula \texttt{$=$ (smoothed map $-$ mean of smoothed off-stream field) / standard deviation of smoothed off-stream field} \citep{yang2023}. The right panel of Figure \ref{fig:SpatialDistribution} shows the calculated S/N across the field. The previously-known satellites and substructures are all clearly detected with high S/N. The red contour indicates where the ${\rm S/N}=15$, a value which captures the most prominent regions of the new S-stream. Specifically,  within this contour, there are 15 times more stars than in the fluctuations in the surrounding region.


\section{PHOTOMETRIC PROPERTIES}\label{section:properties}

\begin{deluxetable*}{lLL}
\tablecaption{Photometric and structural properties of the M83 streams\label{table:properties}}
\tablewidth{0pt} 
\setlength{\tabcolsep}{30pt}
\tablehead{
\colhead{Property} & \colhead{S-stream} & \colhead{N-stream}
}
\startdata
$\alpha$ & 13^{\rm h}36^{\rm m}25\fs65 & 13^{\rm h}37^{\rm m}01\fs50\\
$\delta$ & -30^{\circ}25'03\farcs76 & -29^{\circ}33'27\farcs53\\
projected separation [degree (kpc)] & 0.57~(46) & 0.23~(19)\\
width [degree (kpc)] & 0.11~(9) & 0.10~(8)\\
length [degree (kpc)] & 0.99~(81) & 0.40~(33)\\
$M_{\it V}$ [mag] & -10.72_{-0.54}^{+0.54}~~~\tablenotemark{{\rm (a)}} & -11.34_{-0.49}^{+0.49}~~~\tablenotemark{{\rm (a)}}\\
$M_{\it i}$ [mag] & -11.66_{-0.06}^{+0.06}~~~\tablenotemark{{\rm (a)}} & -12.31_{-0.06}^{+0.06}~~~\tablenotemark{{\rm (a)}}\\
${\it E(B-V)}$ [mag] & 0.05 & 0.04\\
$\langle ({\it g}-{\it i})_0 \rangle$ [mag] & 1.59_{-0.35}^{+0.35} & 1.65_{-0.31}^{+0.31}\\
$\langle [\mathrm{M/H]} \rangle$ [dex] & -1.23_{-0.02}^{+0.02} & -1.05_{-0.01}^{+0.01}\\
$\sigma_{[\mathrm{M/H]}}$ [dex] & 0.28_{-0.01}^{+0.01} & 0.40_{-0.01}^{+0.01}\\
${\rm M}_{*,{\rm flux}}/{\rm M}_{\odot}$ & (8.5_{-2.8}^{+4.2}) \times 10^6~~~\tablenotemark{{\rm (b)}} & (1.6_{-0.5}^{+0.6}) \times 10^7~~~\tablenotemark{{\rm (b)}}\\
${\rm M}_{*,[{\rm Fe/H}]}/{\rm M}_{\odot}$ ($[{\rm \alpha/Fe}]~=~0.0$) & (3.4_{-1.6}^{+3.8}) \times 10^7~~~\tablenotemark{{\rm (c)}} & (13.6_{-6.6}^{+15.5}) \times 10^7~~~\tablenotemark{{\rm (c)}}\\
${\rm M}_{*,[{\rm Fe/H}]}/{\rm M}_{\odot}$ ($[{\rm \alpha/Fe}]~=+0.2$) & (1.2_{-0.5}^{+1.1}) \times 10^7~~~\tablenotemark{{\rm (c)}} & (4.7_{-3.6}^{+5.4}) \times 10^7~~~\tablenotemark{{\rm (c)}}\\
$\langle \mu_{\it V} \rangle$ [mag arcsec$^{-2}$] & 31.8_{-1.9}^{+1.3} & 29.9_{-2.1}^{+1.3}\\
\enddata
\tablecomments{The equatorial coordinates indicate the southernmost position for the S-stream (cyan `x'-mark in the right panel of Fig.~\ref{fig:SpatialDistribution}) and the northernmost position for the N-stream (cyan `+'-mark).}
\tablenotetext{a}{${\it V}$- and ${\it i}$-band absolute magnitudes are calculated under the assumption of the distance to M83.}
\tablenotetext{b}{Estimated from the total flux within the solid red/orange box in the middle panel of Fig.~\ref{fig:SpatialDistribution}.}
\tablenotetext{c}{Estimated from the mass–metallicity relation assuming two $\alpha$-enhancement cases ($[\alpha/\mathrm{Fe}]=0.0$ and $+0.2$).}
\end{deluxetable*}


\subsection{Stellar Populations}\label{subsection:populations}

\begin{figure*}
 \begin{center}
  \includegraphics[width=180mm]{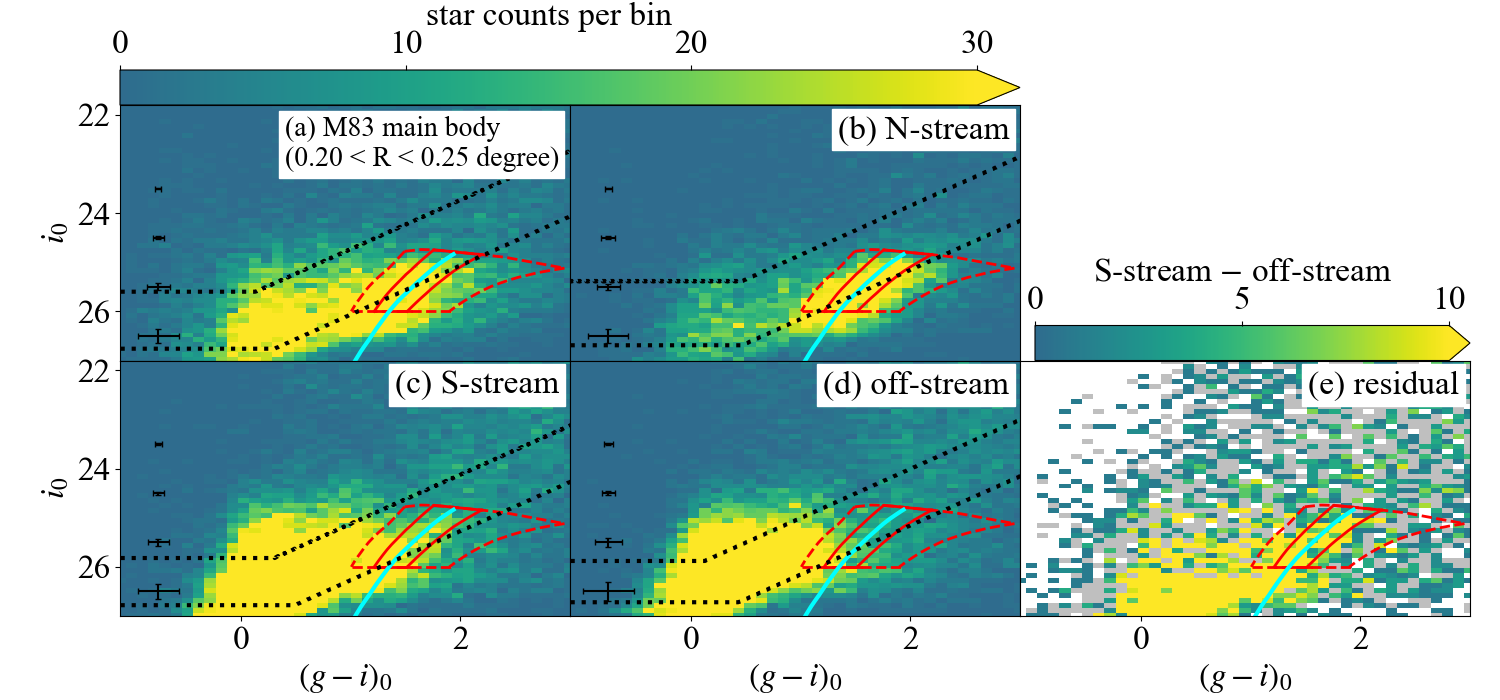}
 \end{center}
 \caption{De-reddened Hess diagrams of the point sources in M83, the N-stream field, the S-stream field, the off-stream field, and the field-subtracted S-stream. The bin sizes are set to 0.1 mag along both the x- and y-axes. Panel (a): Stars located between 0.20 degree and 0.25 degree from the center of M83. The solid (dashed) red box indicates the narrow (wide) RGB box used to select M83 RGB stars. The cyan solid line indicates the PARSEC isochrone \citep{bressan2012} of age 10 Gyr and $[{\rm M/H}] = -1.23$ shifted to M83's distance. This cyan line illustrates the mean photometric metallicity of the S-stream estimated in Section \ref{subsection:metallicity}. Error bars show the median photometric uncertainties of observed sources for each region. The black dashed curves represent the detection completeness limits at the 50\% and 20\% levels. Panel (b), (c), (d): The same as in panel (a), but for the stars within the N-stream polygon, the S-stream polygon, and the off-stream field. Panel (e): The field-subtracted Hess diagram of the S-stream field, constructed by subtracting panel (d) from (c). Gray-colored bins show those in which the count has negative values}\label{fig:CMD}
\end{figure*}

\begin{figure*}[!ht]
 \begin{center}
  \includegraphics[width=175mm]{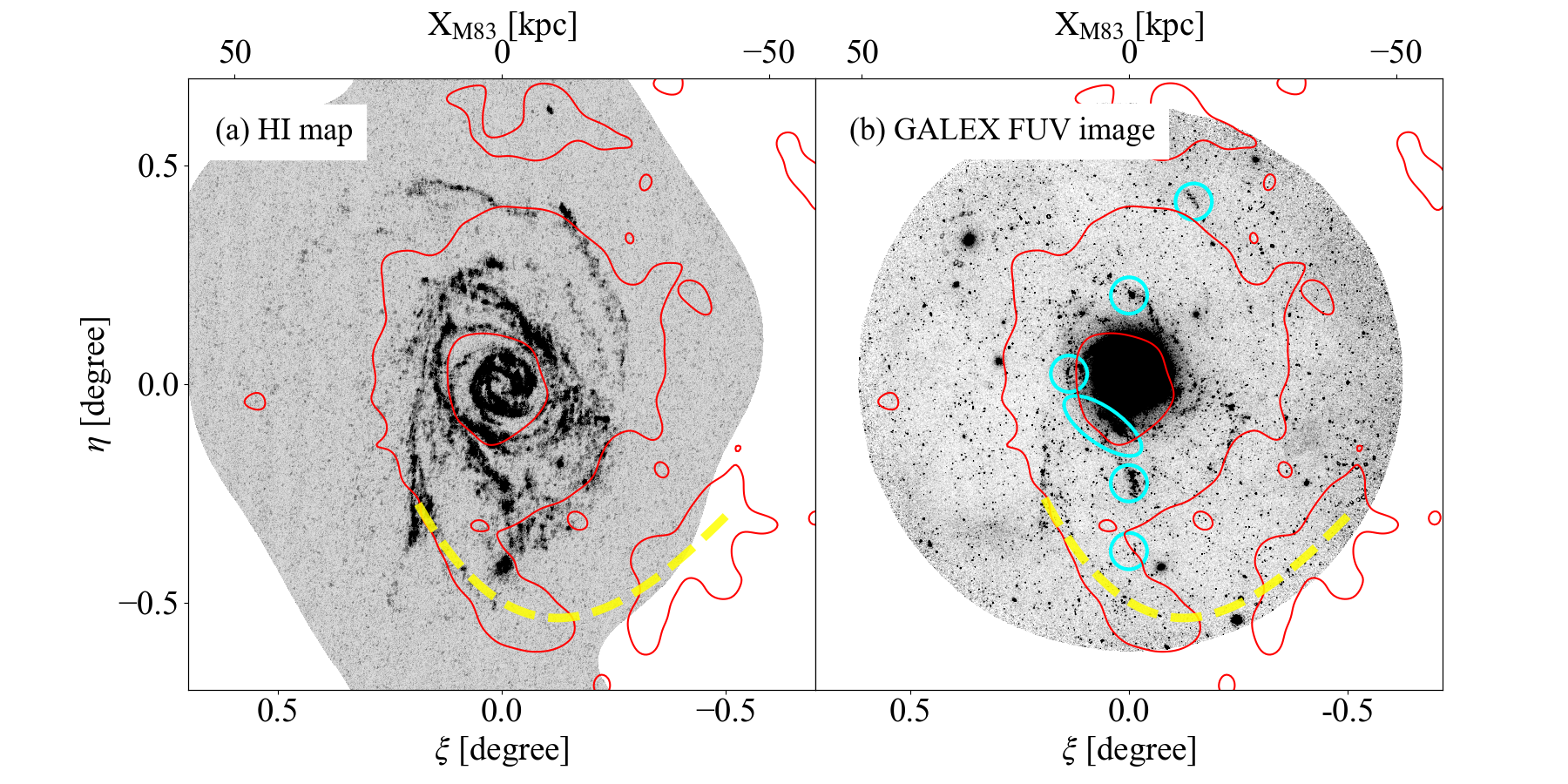}
 \end{center}
 \caption{Panel (a): The \HI map from \citet{eibensteiner2023}. The red contour shows the region within which the S/N of the binned smoothed RGB star-count map is $\geq15$ (see right panel of Figure \ref{fig:SpatialDistribution}), while the yellow dashed line represents the ridge line of the S-stream. Panel (b): The GALEX FUV image. Cyan circles and ellipses are regions that show peaks in the FUV emission defined by \citet{eibensteiner2023}.
 Red contour and yellow dashed line are the same as in Panel (a). }\label{fig:StellarPopulations}
\end{figure*}

Figure \ref{fig:CMD} shows the de-reddened Hess diagrams of the point sources around M83, N-stream, S-stream, off-stream, and the difference between S-stream and off-stream. The Hess diagram of M83 is constructed using stars located between 0.20 degree and 0.25 degree from the center of M83 which is indicated as the cyan solid/dashed circles in the middle panel of Figure \ref{fig:SpatialDistribution}, and those of N-stream, S-stream, and off-stream are constructed using stars in the orange and red polygons in the middle panel of Figure \ref{fig:SpatialDistribution}. Error bars in panels (a) to (d) show the median photometric uncertainties (from hscPipe) of observed sources in the interval $-0.5<({\it g}-{\it i})_0<3.0$ for each region. The cyan lines in all panels correspond to a PARSEC isochrone \citep{bressan2012} with the mean metallicity of the S-stream ($[{\rm M/H}] = -1.23$, estimated in Section \ref{subsection:metallicity}), assuming a distance of 4.7 Mpc to M83 and age of 10 Gyr. The black dashed curves indicate the 50 \% and 20 \% detection completeness limits for each region. In Figure \ref{fig:CMD}, the solid red polygon encloses the RGB stars associated with the newly identified S-stream, while the dashed red polygon represents a wider RGB region that encompasses both metal-poor and metal-rich RGB stars. The wide RGB box is used in constructing the metallicity distribution presented in Section \ref{subsection:metallicity}.

In panels (a) and (b), clear broad RGB sequences are seen from $((g-i)_0,~i_0)\sim(2,~24.5)$ to $(1.5,~26)$ for the outer disk and N-stream. Comparing panels (c) and (d),  a narrower RGB sequence can be seen which aligns with $[{\rm M/H}]=-1.23$ isochrone in the S-stream field, but no such sequence is seen in the off-stream field. To enhance this feature, the difference between the S-stream and off-stream Hess diagrams is shown in panel (e) of Figure \ref{fig:CMD}. Net observed counts within the narrow RGB box are clearly positive, indicating a genuine enhancement compared to the off-stream field. In panel (e) of Figure \ref{fig:CMD}, an overdensity can be seen around $((g - i)_0,~i_0) \sim (0,~25.5)$. These objects have also been reported in previous deep observations of nearby galaxies \citep[e.g.,][]{barker2009,okamoto2015} and are considered to be unresolved background sources. 

Figure \ref{fig:StellarPopulations} compares the morphology of the stream with that of the young stars and \HI gas in M83. Panel (a) shows the \HI map from \cite{eibensteiner2023}, while panel (b) shows the GALEX far-UV (FUV) image. In both panels, we overlay the S/N $= 15$ boundary from the binned star-count map, and the ridge line of the S-stream. 
In panel (b), the cyan circles and ellipse correspond to regions that show peaks in the FUV emission defined by \citet{eibensteiner2023}. It can be seen that the S-stream appears to partially trace the outer \HI and FUV distributions. In panel (a), the western base of the S-stream seems to overlap with the tail of the asymmetric \HI distribution, suggesting that they might be related. Furthermore, the FUV image (panel (b)) exhibits a slight enhancement in brightness at this position, although the brightest FUV peaks do not lie near the stream ridge line. Nevertheless, this general coincidence of features suggests that they could all be linked via a past merger event that disturbed the existing outer gas disk, triggering the star formation that led to the formation of the XUV disk. 

\subsection{Photometric Metallicity and Luminosity}\label{subsection:metallicity}

\begin{figure}
 \begin{center}
  \includegraphics[width=83mm]{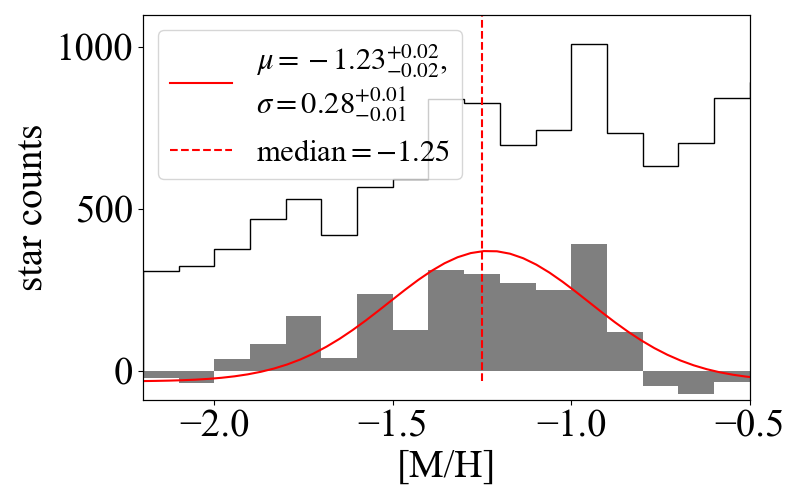}
 \end{center}
 \vspace{-5mm}
 \caption{Completeness-corrected photometric metallicity distribution of RGB stars in the S-stream field. The open histogram shows the metallicity distribution of RGB stars, selected using the wide RGB-box, in the S-stream field, while the gray histogram shows the contaminant-subtracted metallicity distribution using the off-stream field. The solid red line shows a Gaussian fit to the contaminant-subtracted metallicity distribution, and the dashed red line illustrates the median value.}\label{fig:MD}
\end{figure}

To investigate the nature of the S-stream, we estimate the photometric metallicity for individual RGB stars. To do so, we perform Radial Basis function interpolation between PARSEC isochrones with $-2.2<[{\rm M/H}]<-0.5$ (assuming an age of 10 Gyr, which is identical to that in \citet{bell2026}) to convert stellar colors into metallicities. We use stars within the wide RGB box shown in Figure \ref{fig:CMD}, correcting each star for completeness based on its color and magnitude.

Figure \ref{fig:MD} shows the photometric metallicity distribution derived from the completeness-corrected color distribution of RGB stars.  The open histogram corresponds to the S-stream field, while the gray histogram represents the contaminant-subtracted metallicity distribution, which is obtained by subtracting the metallicity distribution of the off-stream field from that of the S-stream field.  The solid red curve represents the result of Gaussian fitting to the contaminant-subtracted metallicity distribution, and the dashed red line indicates the median value of this distribution. To estimate the mean value and standard deviation of the distribution, we invoke the following Bayesian procedure using  Markov Chain Monte Carlo (MCMC) fitting. Assuming that the metallicity distribution follows a Gaussian profile, we construct the model metallicity distribution function as follows:
\begin{equation}
\begin{split}
N_{{\rm model},i} =  \mathcal{C} \times \frac{1}{\sqrt{2\pi\sigma_{\rm MD}^2}} \exp{\left( -\frac{({\rm [M/H]}_{i}-\mu_{\rm MD})^2}{2\sigma_{\rm MD}^2}\right)} \\ + N_{{\rm ref},i},
\end{split}
\end{equation}
where ${\rm [M/H]}_{i}$ denotes a given metallicity value of the observed (binned) metallicity distribution, and $N_{{\rm model},i}$ represents the star counts with a given metallicity ${\rm [M/H]}_{i}$ in the S-stream region. In this analysis, ${\rm [M/H]}_{i}$ refers to a set of metallicity bins spanning the range $-2.2 \leq [{\rm M/H}]_{i} \leq -0.5$, with an interval of 0.1 dex. The model parameters are $\mu_{\rm MD}$, $\sigma_{\rm MD}$, $\mathcal{C}$, which correspond to the mean metallicity, standard deviation, and a normalization constant, respectively. $N_{{\rm ref}, i}$ is the star counts with a given metallicity $[{\rm M/H}]_i$ in the off-stream field. Using this model, we construct the log-likelihood function,
\begin{equation}
\begin{split}
\ln{\mathcal{L}} =  -\frac{1}{2} \sum_i^n \left[ \frac{1}{\sigma_{{\rm obs},i}^2} \left\{ N_{{\rm obs},i} -  N_{{\rm model},i} \right\}^2 + \ln{\sigma_{{\rm obs},i}^2} \right],
\end{split}
\end{equation}
where $N_{{\rm obs}, i}$ and $\sigma_{{\rm obs}, i}$ represent the star counts and the associated Poisson error of the observed metallicity distribution, respectively.
Based on this likelihood function, we perform an MCMC fitting, set to 100 walkers, 110,000 iterations, and first 100,000 burn-in using the Python module \texttt{emcee}. This returns a weighted mean metallicity of $-1.23_{-0.02}^{+0.02}$ dex, and a standard deviation of $0.28_{-0.01}^{+0.01}$ dex. These values and uncertainties correspond to the median and the 68 \% Bayesian credible interval of the posterior distribution. 
Although Figure \ref{fig:MD} exhibits a modest peak, its shape deviates somewhat from a Gaussian profile, raising the possibility that the distribution could be affected by residual contamination from background galaxies. Given that such objects are most likely to contaminate the blue side of the RGB, this would suggest that the mean metallicity could be slightly higher than our estimate. Finally, we estimate the metallicity of the N-stream using the same method. Its metallicity is $\langle[{\rm M/H}]\rangle = -1.05\pm0.01$ dex with a standard deviation of $0.40\pm0.01$ dex, indicating that the N-stream is slightly more metal-rich compared to the S-stream.


In addition to the photometric metallicity, the total luminosity, color, and surface brightness of the S-stream are estimated by analyzing the cumulative flux from the RGB stars, using the same methods as in \citet{zemaitis2023} and \citet{okamoto2024}. The flux of stars fainter than those in the RGB selection box is taken into account using the best-fit isochrone, and the detection completeness of each star is accounted for. The contribution from contamination (foreground stars and background objects) is estimated from the off-stream field and then subtracted. For the surface brightness, we calculate the area enclosed by the red polygon in the middle panel of Figure \ref{fig:SpatialDistribution}. We find that the total $i$-band absolute magnitude is $M_{\it i} = -11.66_{-0.06}^{+0.06}$ and $(g - i)_0 = 1.59_{-0.35}^{+0.35}$, which translates to $M_{\it V} = -10.72_{-0.54}^{+0.54}$ based on the transformations from \citet{komiyama2018}. We find the average surface brightness of the stream to be $\langle \mu_{\it V} \rangle = 31.8_{-1.9}^{+1.3}$. Finally, using the integrated magnitude and color, we use Equation (8) from \citet{taylor2011} to calculate a stellar mass of $(8.5_{-2.8}^{+4.2}) \times 10^6~{\rm M_{\odot}}$.
Applying the same procedures to the N-stream, we obtain $M_{\it i} = -12.31_{-0.06}^{+0.06}$ ($M_{\it V} = -11.34_{-0.49}^{+0.49}$), $(g-i)_0 = 1.65_{-0.31}^{+0.31}$, ${\rm M}_{*,\rm flux}/{\rm M}_{\odot} = (1.6_{-0.5}^{+0.6}) \times 10^7$, and $\langle \mu_{\it V} \rangle = 29.9_{-2.1}^{+1.3}$. 

We can also use the mean photometric metallicity ($[{\rm M/H}] = -1.23$) along with the mass-metallicity relation \citep{kirby2013} to estimate the stellar mass of the progenitor of the S-stream. To convert our measured [M/H] value to [Fe/H], we use Equation (3) of \citet{salaris1993} and explore alpha-element abundances of $[{\rm \alpha/Fe}]=+0.2$ and $[{\rm \alpha/Fe}]=0.0$. This yields $[{\rm Fe/H}]=-1.37$ ($[{\rm Fe/H}]=-1.23$) for $[{\rm \alpha/Fe}] =+0.2$ ($[{\rm \alpha/Fe}] = 0.0$). If the progenitor of the S-stream can be approximated as a single stellar population, its original stellar mass is estimated to be $(1.2_{-0.5}^{+1.1}) \times 10^7~{\rm M}_{\odot}$ for the case of $[{\rm Fe/H}]=-1.37$ ($[\rm \alpha/Fe] = +0.2$) or $(3.4_{-1.6}^{+3.8}) \times 10^7~{\rm M}_{\odot}$ for the case of $[{\rm Fe/H}]=-1.23$ ($[\rm \alpha/Fe] = 0.0$). These values are 1.4 -- 4 times higher than the mass presently contained within the S-stream, albeit with large uncertainties. This discrepancy exists even if we include the N-stream in the calculation. The metallicity of the N-stream implies a stellar mass from the mass-metallicity relation of $(4.7_{-3.6}^{+5.4}) \times 10^7~{\rm M}_{\odot}$ for the case of $[{\rm \alpha/Fe}] =+0.2$, or $(13.6_{-6.6}^{+15.5}) \times 10^7~{\rm M}_{\odot}$ for the case of $[{\rm \alpha/Fe}] = 0.0$. In both cases, the central estimate inferred by the mass-metallicity relation exceeds the combined photometric stellar mass of the two streams, $(2.5_{-3.3}^{+4.8}) \times 10^7~{\rm M}_{\odot}$. This difference could indicate additional undetected debris.

\section{DISCUSSION AND CONCLUSIONS}\label{section:conclusions}


We have studied the faint stellar outskirts of the nearby spiral galaxy M83 using data obtained from Subaru/HSC. The depth of the data ($i_0\sim 26$ mag) allows us to construct CMDs that reach $\sim 2$ magnitudes below the TRGB, and construct a map of resolved RGB stars.  In addition to recovering previously-known features such as two satellite galaxies and the prominent N-stream, we also confirm the recently-reported new southern stream (S-stream). 

Our deep wide-area map allows us to trace the S-stream for $\sim 81$ kpc along its length, with a relatively large width of $\sim 9$ kpc.
The projected RGB distribution also shows apparent density variations along the S-stream (Figure \ref{fig:StellarPopulations}). We do not interpret these features as evidence for encounters with dark matter subhalos \citep[e.g.,][]{yoon2011}, nor does their appearance establish an origin in the progenitor's internal structure. Assessing their significance requires accounting for spatial variations in completeness, contamination, and the selection function. If confirmed, their physical origin would require further investigation, including the possible roles of variations in the stripping rate, orbital phase, projection effects, and the progenitor's stellar distribution.

The inferred photometric properties indicate that the S-stream is composed of old, fairly metal-poor ([M/H]$=-1.23$) stars.  The N-stream is slightly more metal-rich ([M/H]$=-1.05$) and higher surface brightness than the S-stream, but our wide-area map confirms the suggestion by \citet{bell2026} that these two features are very likely related and originate from a single minor merger. Based on the higher metallicity ([M/H]$=-1.05$) of the N-stream, the mass-metallicity relationship predicts that the progenitor has a stellar mass of $\left( 4.7_{-3.6}^{+5.4} \right) \times10^7~{\rm M}_{\odot}$ for the case of $[{\rm \alpha/Fe}]=+0.2$ and $\left( 13.6_{-6.6}^{+15.5} \right) \times10^7~{\rm M}_{\odot}$ for the case of $[{\rm \alpha/Fe}]=0.0$.  Although the uncertainties are considerable, this value is somewhat higher than the total observed mass of the two streams, implying that there may be other debris from the progenitor elsewhere in the halo.

To place the inferred progenitor mass in the context of the merger mass ratio, we adopt the stellar mass of M83 derived from WISE photometry by \citet{heald2016}, rescaled to our adopted distance of 4.7 Mpc, giving ${\rm M_{\star,\mathrm{M83}}}\simeq2.9\times10^{10}~{\rm M_\odot}$. Their estimate uses the calibration of \citet{cluver2014}, which is tied to the GAMA stellar masses of \citet{taylor2011} and is therefore consistent with the IMF normalization adopted for our stream mass estimates. Assuming that the S- and N-streams originate from a single progenitor, their combined photometric stellar mass gives
$\frac{M_{\star,S}+M_{\star,N}}{M_{\star,\mathrm{M83}}}
\simeq8.4\times10^{-4}$,
or approximately $1:1200$. This is the ratio of the stellar mass associated with the detected debris to the present-day stellar mass of M83; it does not account for debris outside the measured regions or any surviving remnant.
Using instead the progenitor stellar masses inferred above from the N-stream metallicity gives approximate progenitor-to-host stellar mass ratios of $1:620$ and $1:210$ for $[\alpha/\mathrm{Fe}]=+0.2$ and $0.0$, respectively. These values are illustrative central estimates rather than bounds on the mass ratio. They depend on the applicability of the stellar mass--metallicity relation and on how well the N-stream metallicity represents the original progenitor. Nevertheless, they support an accretion event involving a satellite with a small stellar mass relative to M83. These estimates use the present-day host stellar mass and do not directly determine the mass ratio at infall. Nor do they constrain the progenitor's halo mass at infall or its remaining bound mass during subsequent disk encounters. The inferred stellar mass ratios alone therefore do not establish the strength of the gravitational perturbation to M83's outer gas disk.

We compare our stream measurements with those reported by \citet{bell2026}. Their stellar mass estimates are $1.45^{+0.37}_{-0.30}\times10^7\,M_\odot$ for the S-stream and $7.08_{-1.33}^{+1.63}\times10^7\,M_\odot$ for the N-stream. These central values exceed ours by factors of approximately 1.7 and 4.4, respectively. The quoted uncertainty intervals overlap for the S-stream but not for the N-stream, indicating that the latter difference warrants particular attention.

Both studies find that the N-stream is approximately $0.2$~dex more metal-rich than the S-stream. Our quoted mean metallicities are obtained from Gaussian fits to the corrected metallicity distributions, whereas \citet{bell2026} report medians. Although these estimators need not yield identical values for non-Gaussian distributions, both analyses recover a similar metallicity difference between the two streams.

The metallicity difference does not necessarily preclude a common progenitor. If the progenitor had a negative radial metallicity gradient, tidal stripping could preferentially remove its less tightly bound, more metal-poor outer stars before exposing its more metal-rich inner populations. Different portions of the resulting debris could therefore have different mean metallicities if they sample different mixtures of these populations and stripping epochs. Such a mechanism has been explored in models of the Sagittarius stream \citep[e.g.,][]{law2010}. In this framework, the lower metallicity of the S-stream could reflect a larger contribution from the progenitor's outer population, whereas the N-stream could contain a larger contribution from more centrally concentrated stars.

Separate progenitors remain possible, including progenitors accreted as members of the same satellite group. In the Auriga simulations, \citet{riley2025} identify streams produced by satellites disrupting around more massive satellites, illustrating the role of preprocessing in stream formation. This provides a possible alternative framework in which the S- and N-streams could share a related accretion history without originating from the same galaxy, and their metallicity difference would not require an internal gradient in a single progenitor. However, the present data do not establish such a group-infall origin.

As described in Section~\ref{section:data}, the additional M83-centered pointing improves the imaging depth particularly in the eastern region of M83. This provides greater sensitivity to faint structure in that region, but the measured luminosity and mean surface brightness also depend on the adopted stream footprint. Integrated luminosity measures the light enclosed within that footprint, whereas mean surface brightness additionally depends on its area. Measurements over different regions therefore need not yield identical values, even if they sample the same physical stream. Furthermore, because the studies share part of the archival HSC imaging, their measurements are not fully observationally independent.

The luminosity and stellar mass estimators also differ. We estimate the total luminosity from the completeness-corrected, contamination-subtracted RGB flux, including an isochrone-based correction for unobserved stars, and apply the color-dependent mass-to-light relation of \citet{taylor2011}. By contrast, \citet{bell2026} estimate luminosities and masses using RGB star-count scalings. Differences in source selection, background subtraction, completeness corrections, and stellar-population conversions can therefore affect the comparison. These observational and methodological differences provide possible contributions to the reported offsets.

We interpret the S-stream as stellar debris from a disrupted satellite, based on its extended, coherent morphology and metal-poor RGB population. Its possible connection with the N-stream also motivates a common accretion origin. In this interpretation, the partial overlap with the outer \HI structures could arise if the stellar debris is accompanied by gas stripped from the progenitor, or if the interaction has perturbed and redistributed gas already present in M83. These possibilities are not mutually exclusive, and the progenitor need not have supplied the bulk of the extended \HI disk.

An alternative possibility is that the stellar feature is a tidal tail extracted from M83's own disk. Galaxy encounters can draw disk material into elongated tidal structures \citep[e.g.,][]{toomre1972}, and such a feature could contain pre-existing disk stars as well as redistributed disk gas. Neither the stream-like morphology nor the presence of RGB stars therefore uniquely establishes a satellite origin. Although the metal-poor stellar population is consistent with disrupted dwarf-galaxy debris, our photometric metallicities alone cannot exclude an origin in M83's outer stellar disk. The oxygen abundances measured in present-day outer-disk H II regions \citep{bresolin2009} are not directly comparable to the metallicities of the older stars that could have been stripped. Likewise, the \HI kinematic disturbances discussed by \citet{heald2016} provide evidence for perturbations to the outer gas disk but do not determine the origin of the stream stars. The spiral-like appearance of the \HI structures and their projected overlap with the stream are therefore insufficient to distinguish these scenarios, and a chance superposition cannot be excluded. We retain satellite disruption as our working interpretation and consider below how this accretion event could have affected M83's pre-existing gas disk and contributed to its XUV star formation. Stellar kinematics and encounter models would be needed to distinguish this interpretation from a disk-derived tidal tail.

The stream identified in this study shows some correspondence with features in the \HI and FUV maps (Figure \ref{fig:StellarPopulations}), suggesting a possible connection between the accretion event that produced the stream and the formation of M83's XUV disk. If the progenitor was accreted on a low-inclination eccentric orbit, it could have interacted with M83's outer \HI disk, redistributing and compressing existing gas. Together with any gas supplied by the progenitor, these disturbances could have triggered star formation and contributed to the development of the XUV disk.
Evidence for disturbances in the outer gas disk comes from the systematic distortions in the \HI isovelocity contours coinciding with the N-stream reported by \citet{heald2016}, which they interpreted as evidence of a gravitational influence on the outer-disk gas. These disturbances are consistent with a tidal interaction, but do not distinguish gas supplied by the satellite from pre-existing M83 gas perturbed during the encounter.

Redistribution and mixing of chemically enriched gas from smaller radii would also be consistent with the relatively high oxygen abundance and flat radial abundance profile measured in outer-disk \HII regions \citep{bresolin2009}.
If the common progenitor of the S- and N-streams supplied a substantial amount of gas with a lower oxygen abundance than the receiving disk gas, mixing would tend to dilute the local oxygen abundance. The magnitude and spatial extent of this effect would depend on the amount and composition of the accreted gas and the subsequent mixing. The published \HII-region abundances do not by themselves determine this contribution, and our imaging provides no direct measurement of a chemical anomaly specifically associated with the stream. Moreover, the metallicities of the stream RGB stars do not directly specify the oxygen abundance of the progenitor's gas at accretion. Thus, the outer-disk abundance profile is compatible with redistribution of enriched gas, but neither establishes nor excludes an additional contribution from satellite gas.
The high-velocity molecular clouds detected by \citet{nagata2025}, most of which they interpreted as inflowing from outside the galactic disk, provide additional motivation to investigate gas accretion and transport in M83, although their connection to the stream-forming event remains uncertain.

The proposed connection does not necessarily require the progenitor to have supplied the bulk of M83's extended \HI reservoir. Gravitational perturbations alone can enhance star formation in gas already present in a host galaxy, as demonstrated by simulations of gas-rich dwarf galaxies interacting with satellites composed solely of dark matter \citep{starkenburg2016}. Although those simulations cannot be applied quantitatively to M83 without further modelling, they illustrate a mechanism by which an accreting satellite could promote star formation even if it supplied little gas. Gas delivery and gravitational perturbations are not mutually exclusive, and the observed \HI kinematic distortions do not by themselves determine their relative contributions. While \citet{bell2026} argued against the stream-forming accretion event as the source of M83's massive extended \HI disk and the trigger of its nuclear starburst, the possibility considered here is that the interaction reorganized an existing gas reservoir and helped establish the conditions for star formation in the XUV disk.

Our additional \HI and FUV comparison primarily informs the possible response of the outer disk and does not provide a new quantitative constraint on the origin of the nuclear starburst. The reported outer-disk \HI disturbances and the partial morphological correspondence with the stellar stream are consistent with an interaction, but do not demonstrate net gas transport into the nuclear region. Establishing such a connection would require constraints on both the inward gas transport and the timing of the encounter relative to the nuclear starburst. Our analysis does not measure the nuclear inflow rate or gas transport timescale, and the stellar age adopted for the RGB metallicity estimates does not date the accretion event. The present results therefore do not independently establish or exclude a connection to the nuclear starburst, and do not contradict the interpretation of \citet{bell2026}. A contribution to localized star formation in the XUV disk remains possible without requiring the same interaction to have triggered the nuclear starburst.

We therefore consider the minor merger traced by the stellar stream a plausible contributor to the formation of M83's XUV disk. This remains a hypothesis: the projected correspondence is partial, the brightest FUV peaks do not closely follow the stream ridge, and neither the causal connection nor the extent of merger-induced star formation is established by the present data. Spatially resolved abundance measurements targeting the gas structures projected near the streams, together with dynamical and hydrodynamical modelling constrained by the stellar debris and gas kinematics, are needed to distinguish gas delivery from redistribution and to test whether the encounter could reproduce the observed \HI structures and spatial distribution of star formation. Deep imaging of other XUV disks would also help to establish how commonly faint tidal features occur in their outskirts.

It is noteworthy that the surface brightness of the S-stream falls below $\mu_{V} \sim 30$ mag arcsec$^{-2}$, making it one of the lowest surface brightness streams known outside of the Milky Way \citep[see also][]{zemaitis2023}. Although numerous satellite galaxies have been identified around M83 through previous surveys \citep[e.g.,][]{muller2024a}, detections of new substructures have been rare. Owing to their extremely diffuse nature, these features have been challenging to detect in relatively shallow observations \citep[e.g.,][]{muller2015,muller2017}. This highlights the critical importance of surface-brightness depth in searches for faint halo substructures. With the ongoing observations from ESA's Euclid mission and forthcoming data from the Vera C. Rubin Observatory's Legacy Survey of Space and Time, and the Nancy Grace Roman Space Telescope, searches for low surface brightness substructures around large samples of nearby galaxies are becoming increasingly feasible, reaching sensitivities of $\mu>30$ mag arcsec$^{-2}$.

\begin{acknowledgments}
Data analysis was in part carried out on the Multi-wavelength Data Analysis System (MDAS) operated by the Astronomy Data Center (NAOJ/ADC) and the Large-scale data analysis system (LSC) co-operated by the ADC and Subaru Telescope, NAOJ.
Based in part on data collected at Subaru Telescope and obtained from the SMOKA, which is operated by the NAOJ/ADC. 
This work was supported by JSPS Core-to-Core Program (grant number: JPJSCCA20210003), the Overseas Travel Fund for Students (2024) of the Astronomical Science Program, the Graduate University for Advanced Studies, SOKENDAI and the NAOJ Research Coordination Committee, NINS (NAOJ-RCC-2402-0401, NAOJ-RCC-2504-0301).
S.O. acknowledges support in part from JSPS KAKENHI grant Nos. JP18H05875, JP20K04031, JP25K01047. K.H. acknowledges support in part from JSPS KAKENHI grant Nos. JP24K00669, JP25H01553. M.C. acknowledges support from JSPS KAKENHI grant Nos. JP24K00669 and JP25H00394. AMNF is supported by UK Research and Innovation (UKRI) under the UK government’s Horizon Europe funding guarantee [grant number EP/Z534353/1] and by the Science and Technology Facilities Council [grant number ST/Y001281/1]. JK acknowledges support from NSF through grants AST-2006600 and AST-2406608, and from STScI through grant HST-GO-17747.001-A.
The Pan-STARRS1 Surveys (PS1) and the PS1 public science archive have been made possible through contributions by the Institute for Astronomy, the University of Hawaii, the Pan-STARRS Project Office, the Max-Planck Society and its participating institutes, the Max Planck Institute for Astronomy, Heidelberg and the Max Planck Institute for Extraterrestrial Physics, Garching, The Johns Hopkins University, Durham University, the University of Edinburgh, the Queen's University Belfast, the Harvard-Smithsonian Center for Astrophysics, the Las Cumbres Observatory Global Telescope Network Incorporated, the National Central University of Taiwan, the Space Telescope Science Institute, the National Aeronautics and Space Administration under Grant No. NNX08AR22G issued through the Planetary Science Division of the NASA Science Mission Directorate, the National Science Foundation Grant No. AST–1238877, the University of Maryland, Eotvos Lorand University (ELTE), the Los Alamos National Laboratory, and the Gordon and Betty Moore Foundation. 
This work is based in part on observations made with the Galaxy Evolution Explorer (GALEX). GALEX is a NASA Small Explorer, whose mission was developed in cooperation with the Centre National d'Etudes Spatiales (CNES) of France and the Korean Ministry of Science and Technology. GALEX is operated for NASA by the California Institute of Technology under NASA contract NAS598034. 
\end{acknowledgments}

\clearpage
\facilities{Subaru (HSC)}
\software{emcee \citep{foreman-mackey2013},
          astropy \citep{theastropycollaboration2013},
          matplotlib \citep{hunter2007},
          numpy \citep{vanderwalt2011},
          corner \citep{foreman-mackey2016},
          dustmaps \citep{m.green2018},
          hscPipe \citep{bosch2018}}

\appendix
\section{STAR-GALAXY CLASSIFICATION BASED ON A DETERMINANT RADIUS}\label{section:detR}

This section introduces our method of star-galaxy classification, based on the determinant radius, $R_{\rm det}$. The determinant radius is a parameter relating to the size of an object. It is defined in \citet{aihara2018} as:
\begin{equation}
R_{\rm det} = ({\rm shape}_{\rm xx}~\times~{\rm shape}_{\rm yy} – {\rm shape}_{\rm xy}^2)^{1/4},\label{eq:detR}
\end{equation}
where ${\rm shape_{xx}}$, ${\rm shape_{yy}}$, and ${\rm shape_{xy}}$ are the Gaussian-weighted second-order adaptive moments. These moments are computed individually for each filter and each object. Additionally, the moments of the PSF model at the position of each object are calculated. Using these quantities and Equation (\ref{eq:detR}), we derive the determinant radius of the elliptical Gaussian which corresponds to an object's radius ($R_{\rm det, obj}$), and the PSF model's determinant radius ($R_{\rm det, psf}$) at the corresponding position. Their difference, $\Delta R_{\rm det} \equiv R_{\rm det, obj}-R_{\rm det, psf}$, is then calculated for each object.

\begin{figure}
 \begin{center}
 \includegraphics[width=0.8\linewidth]{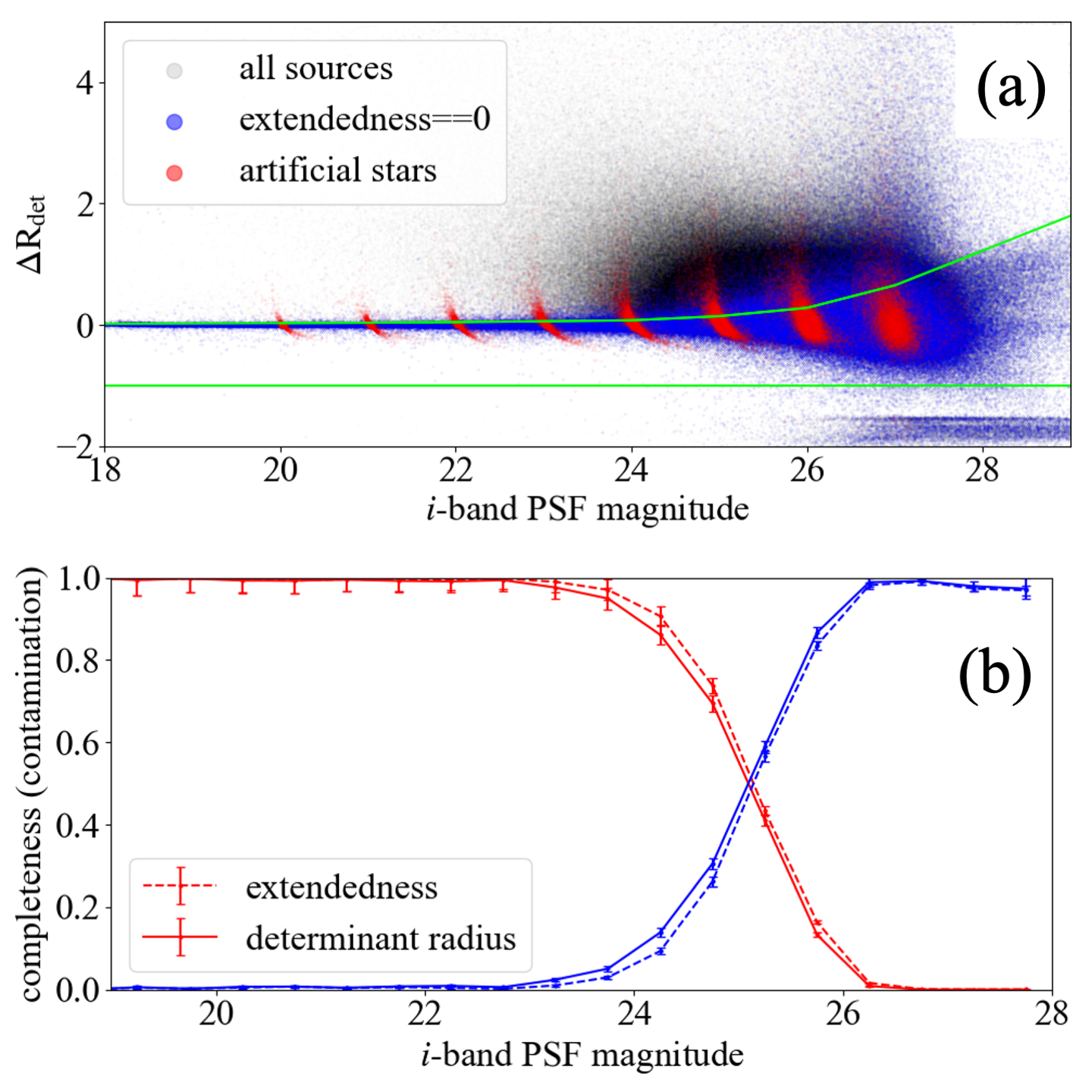}
 \end{center}
 \caption{Panel (a): Difference in determinant radius $\Delta R_{\rm det}$ as a function of {\it i}-band PSF magnitude. Black dots are all detected objects, blue dots illustrate objects with $\texttt{extendedness} == 0$, and red dots represent recovered artificial stars with magnitudes ranging from 20 to 28 mag in 1 mag intervals. The light-green curves are boundaries used to select the point sources in this study. The upper one shows the 3 times standard deviation of $\Delta R_{\rm det}$ for the artificial stars calculated by the DBSCAN method, and the lower one is a boundary defined to cut artifacts and/or noise. Panel (b): Completeness (red) and contamination (blue) of the star-galaxy classification as a function of the {\it i}-band PSF magnitude. Dashed lines represent the results obtained using the \texttt{extendedness} parameter, while solid lines represent the results based on the determinant radius.}\label{fig:detR_AST}
\end{figure}

The panel (a) in Figure \ref{fig:detR_AST} illustrates the difference in determinant radius, $\Delta R_{\rm det}$, as a function of {\it i}-band PSF magnitude. In this figure, the black dots represent all detected objects, the blue dots correspond to objects with $\texttt{extendedness} == 0$, and the red dots indicate artificial stars with magnitudes ranging from 20 to 28 mag, in 1 mag intervals. In this plot, point sources, which have no shape difference compared to the PSF model, are located around $\Delta R_{\rm det} \sim 0$, while extended sources such as galaxies are distributed at $\Delta R_{\rm det} > 0$.

In this study, we construct a point-source catalog based on the distribution of recovered artificial stars, shown as red dots in panel (a) of Figure \ref{fig:detR_AST}. For each magnitude bin, we model the distribution of $\Delta R_{\rm det}$ for the artificial stars as a function of PSF magnitudes. To achieve this, we use DBSCAN to obtain the deviation of the artificial stars for each magnitude. DBSCAN is a density-based clustering algorithm that identifies groups of data points as regions of high point density separated by areas of low density \citep{ester1996}. The distribution of artificial stars is non-Gaussian, so outlier-removal techniques based on Gaussian statistics, such as the sigma-clipping method, are not optimal. In panel (a) of Figure \ref{fig:detR_AST}, the upper light-green curve indicates three times the standard deviation calculated from the objects selected by DBSCAN, and we adopt this curve as the boundary between point sources and extended sources. For DBSCAN, we adopt a neighborhood radius equal to three times the typical photometric uncertainty, and set the minimum number of samples for a cluster to 1 \% of the total samples, which is a commonly used value. Objects located around $\Delta R_{\rm det} \sim -2$ are considered artifacts or noise, so we choose the objects with  $\Delta R_{\rm det} > -1$, shown as the lower light-green curve, as the lower boundary. 

To evaluate the validity of our star-galaxy separation based on the determinant radius, we calculate the completeness and contamination of our classification using HSC-SSP data and HST/Advanced Camera for Surveys (ACS) data \citep{leauthaud2007} in the COSMOS field \citep{scoville2007}, following the same methodology as in \citet{aihara2018}. The completeness is defined as the fraction of HST/ACS stars properly classified as stars in the HSC catalog. The contamination is the fraction of HST/ACS galaxies among objects classified as stars in the HSC catalog. Panel (b) of Figure \ref{fig:detR_AST} shows the completeness (red) and contamination (blue) of our star-galaxy classification as a function of the {\it i}-band PSF magnitude. The dashed lines represent the results obtained based on the \texttt{extendedness} parameter, while the solid lines represent the results based on the determinant radius. It can be seen that our classification method based on the determinant radius achieves an accuracy comparable to that obtained based on the \texttt{extendedness} parameter. It should be noted that the HSC data in the COSMOS field used here are of higher quality than the M83 data used in this study. Therefore, panel (b) merely demonstrates that the determinant radius method achieves a similar level of performance as the \texttt{extendedness} parameter and does not represent the completeness (contamination) value of our star-galaxy classification in this study. For instance,  panel (b) of Figure \ref{fig:detR_AST} shows that the completeness to classify point sources at ${\it i} \sim 25$ mag is 50\%.

\newpage
\bibliography{Ogami+2025b_resubmit}{}
\bibliographystyle{aasjournal}

\end{document}